\documentclass[fleqn,12pt,twoside]{article}
\usepackage{espcrc1}

\usepackage{graphicx}
\usepackage[figuresright]{rotating}

\newcommand{\sfrac}[2]{\mbox{\footnotesize $\displaystyle \frac{#1}{#2}$}} 
\newcommand{\nlsim}{\mathrel{\rlap{\lower4pt\hbox{\hskip0pt$\sim$}} 
 \raise1pt\hbox{$<$}}}           
\newcommand{\ngsim}{\mathrel{\rlap{\lower4pt\hbox{\hskip0pt$\sim$}} 
 \raise1pt\hbox{$>$}}}           

\newcommand{\AmS}{{\protect\the\textfont2
  A\kern-.1667em\lower.5ex\hbox{M}\kern-.125emS}}

\title{Theory and Phenomenology of Hadrons\thanks{CDR is grateful to the organisers for their efficiency, hospitality and support during this meeting and in the preceding few days, and acknowledges fruitful conversations with B.~El-Bennich, B.~Loiseau and P.\,C.~Tandy.  This work was supported by the: US Department of Energy, Office of Nuclear Physics, contract no.\ W-31-109-ENG-38; Austrian Science Fund FWF, Schr\"odinger-R\"uckkehrstipendium R50-N08; 
and benefited from the ANL Computing Resource Center's facilities.}}

\author{M.\,S.\ Bhagwat,\address[anl]{Physics Division, Argonne National Laboratory,
Argonne, IL, 60439, USA}
        A.\ H\"oll\address[rostock]{Institut f\"ur Physik, Universit\"at Rostock, D-18051 Rostock, Germany}
        A.\ Krassnigg\address[graz]{Fachbereich Theoretische Physik, Universit\"at Graz, A-8010 Graz, Austria}
        and
        C.\,D.\ Roberts\addressmark[anl]
}
       
\begin{document}

\maketitle

\begin{abstract}
This Dyson-Schwinger equation (DSE) aper\c{c}u highlights recent applications to mesons.  It reports features of, and results for, pseudoscalar and scalar bound-state residues in vacuum polarisations, and exhibits how a restoration of chiral symmetry in meson trajectories could be manifest in a relationship between them.  It also touches on nucleon studies, emphasising the importance of both scalar and axial-vector diquark correlations, and reporting the calculation of $\mu_n\, G_E^n(Q^2)/G_M^n(Q^2)$.  The value of respecting symmetries, including Poincar\'e covariance, is stressed.
\end{abstract}

\section{INTRODUCTION}
``While there is considerable optimism that lattice gauge theory will continue to improve as a means of studying nonperturbative QCD, it is also extremely important to pursue covariant, nonperturbative approximation methods.  In particular, there is a continuing need for the development of approximation techniques and models which bridge the gap between perturbative QCD and the large amount of low- and intermediate-energy phenomenology in a single covariant framework.''  Thus began Ref.\,\cite{Williams:1989tv} and although these observations remain true today, material progress has been made on all fronts.  This contribution provides a glimpse of advances made with DSEs since \emph{Few Body 17}.  

The DSEs are a nonperturbative tool for the study of quantum field theory in general \cite{Roberts:1994dr} and continuum strong QCD in particular. They provide a generating tool for perturbation theory and, since QCD is asymptotically free, this means that any model-dependence in the application of these methods can be restricted to the long-range domain.  In this mode, the DSEs provide a tool with which strong interaction phenomena can be used to map out, e.g., the behaviour at long range of the interaction between light-quarks.  This is not accessible by other means (see, e.g., Refs.\,\cite{Krein:1990sf,Bali:2005bg}).  DSEs enable the study of: hadrons as composites of dressed-quarks and -gluons; the phenomena of confinement and DCSB, and therefrom an articulation of any connection between them.  The solutions of the DSEs are Schwinger functions, from which all cross-sections can be constructed.  Hence they can be used to make predictions for real-world experiments.  One merit in this is that any assumptions employed, or guesses made, can be tested, verified and improved, or rejected in favour of more promising alternatives.  The modern application of these methods is described in Refs.\,\cite{Roberts:2000aa,Alkofer:2000wg,Maris:2003vk,Fischer:2006ub} with a pedagogical overview provided in Refs.\,\cite{Holl:2006ni,BhagwatSchladming}.

\section{MESONS}
A barrier to the ready application of DSEs is the requirement of truncation, which is made necessary by the coupling between equations.  Naturally, the well-known and much-applied weak-coupling expansion reproduces everything accessible in perturbation theory, but that is inadequate to meet hadron physics' nonperturbative challenges.  Thus the nonperturbative, systematic, and symmetry-preserving truncation introduced in Refs.\,\cite{Munczek:1994zz,Bender:1996bb} has been a boon, enabling the proof of exact results in QCD.  

For example, an exact mass formula for flavour-nonsinglet pseudoscalar mesons was derived \cite{Maris:1997hd}: $f_H m_H^2 = [m_1(\zeta)+m_2(\zeta)]\rho_H(\zeta)$, in which $m_{1,2}(\zeta)$ are the running current-quark masses of the constituents, $\zeta$ is the renormalisation mass-scale, and 
\begin{eqnarray}
\label{fH}
f_{H} \, P_\mu &=& Z_2 {\rm tr} \int_q^\Lambda \! \sfrac{1}{2} (T^H)^T \gamma_5
\gamma_\mu {\cal S}(q_+)\, \Gamma_{H}(q;P)\, {\cal S}(q_-)\,,\\
i  \rho_{H}\!(\zeta)\,  &=& Z_4\,{\rm tr} 
\int^\Lambda_q 
(T^H)^T \,  \gamma_5 \,  {\cal S}(q_+)\, \Gamma_{H}(q;P)\, {\cal S}(q_-)\,,
\end{eqnarray}
where: $Z_{2,4}$ are the quark wave-function and Lagrangian-mass renormalisation constants; $\int^\Lambda_q := \int^\Lambda d^4 q/(2\pi)^4$ represents a \textit{Poincar\'e-invariant} regularisation of the integral, with $\Lambda$ the regularisation mass-scale; the matrices $T^H$ are constructed from the generators of $SU(N_f)$ with, e.g., \mbox{$T^{\pi^+}=\mbox{\small $\frac{1}{2}$} (\lambda^1+i\lambda^2)$} providing for the flavour content of a positively charged pion; $q_\pm = q\pm P/2$; ${\cal S}= {\rm diag}[S_u,S_d,S_s,\ldots]$; and $\Gamma_{H}(q;P)$ is the meson's Bethe-Salpeter amplitude.  The Gell-Mann--Oakes--Renner relation is a corollary.  In deriving these expressions, reference is not made to the magnitude of the current-quark mass.  Hence they apply equally to heavy-light systems, in which case follows \cite{Ivanov:1997iu} $f_H \propto 1/\hat m_Q$ and \cite{Ivanov:1998ms}
$m_H \propto \hat m_Q \;\; \mbox{for} \;\; 1/\hat m_Q\sim 0\,,$
where $ \hat m_Q$ is the renormalisation-group-invariant current-quark mass of the meson's heaviest constituent.  Moreover, the mass formula applies not just to ground states but to all mesons on the $J^{PC}=0^{-+}$ trajectory, whether they are radial excitations or hybrids.  A corollary of this observation is that, as a necessary consequence of chiral symmetry and its dynamical breaking, \cite{Holl:2004fr}
\begin{equation}
\label{fpinzero}
f_{\pi_n}^{\hat m =0} \equiv 0\,, \forall \, n\geq 1\,,
\end{equation}
where $n$ labels a state on the $0^{-+}$ trajectory, with $n=0$ being the ground-state.  Thus, in the presence of dynamical chiral symmetry breaking (DCSB) all pseudoscalar mesons except the ground-state decouple from the weak interaction in the chiral limit.  (At the true current-quark mass $f_{\pi_1}/f_{\pi_0}=-0.02$ \cite{Holl:2004fr}.)  On the other hand, in the absence of DCSB the leptonic decay constant of the ground state $0^{-+}$ also vanishes in the chiral limit, and hence all pseudoscalar mesons are blind to the weak interaction.  The result in Eq.\,(\ref{fpinzero}) can be used as a benchmark to tune and validate lattice-QCD techniques that try to determine the properties of excited state mesons \cite{McNeile:2006qy}.  Exact results for two-photon decays of pseudoscalar mesons are elucidated in Ref.\,\cite{Holl:2005vu}.

Statements may also be made about scalar mesons.  The renormalised Ward-Takahashi identity for the vector vertex can be expressed
\begin{equation}
\label{vwti}
P_\mu i\Gamma_\mu^H(k;P;\zeta) = {\cal S}(k_+) T^H - T^H {\cal S}(k_-) - [{\cal M}^\zeta,T^H] \, i\Gamma_{\mathbf 1}^H(k;P;\zeta)\,,
\end{equation}
where ${\cal M}^\zeta= {\rm diag}[m_u(\zeta),m_d(\zeta),m_s(\zeta),\ldots]$ with, e.g., $m_u(\zeta)$ being the running $u$-quark mass at a renormalisation mass-scale $\zeta$, and $\Gamma_{\mathbf 1}^H(k;P;\zeta)$ is the scalar vertex.  Both $\Gamma_\mu^H(k;P;\zeta)$ and $\Gamma_{\mathbf 1}^H(k;P;\zeta)$ can be obtained from an inhomogeneous Bethe-Salpeter equation.  In the neighbourhood of a scalar bound-state pole
%
\begin{equation}
\left. \Gamma_{\mu,\mathbf 1}^{H}(k;P)\right|_{P^2+m_{\sigma_n^H}^2 \approx 0}= \frac{{\cal R}_{\mu,\mathbf 1}}{P^2 + m_{\sigma_n^H}^2} \Gamma_{\sigma_n^H}(k;P) + \; \Gamma_{\mu,\mathbf 1}^{H\,{\rm reg}}(k;P) \,, 
\left\{\begin{array}{l}
{\cal R}_\mu = f_{\sigma_n^H} \, P_\mu\\
{\cal R}_{\mathbf 1} = \rho_{\sigma_n^H}(\zeta) 
\label{genvv} 
\end{array}\right.;
\end{equation}
viz., each vertex may be expressed as a simple pole plus terms regular in its  neighbourhood, with $\Gamma_{\sigma_n^H}(k;P)$ representing the bound-state's normalised Bethe-Salpeter amplitude, 
%
and
\begin{eqnarray} 
\label{fsigman} f_{\sigma_n^H} \,  P_\mu &=& Z_2\,{\rm tr} \int^\Lambda_q 
(T^H)^T  \gamma_\mu\, {\cal S}(q_+)\, \Gamma_{\sigma_n^H}(q;P)\, {\cal S}(q_-) \,, \\
\label{cpres} \rho_{\sigma_n}\!(\zeta)\,   &=& Z_4\,{\rm tr} 
\int^\Lambda_q (T^H)^T \, {\cal S}(q_+)\, \Gamma_{\sigma_n^H}(q;P)\, {\cal S}(q_-)\,.
\end{eqnarray} 
The residues expressed in Eqs.\,(\ref{fsigman}) and (\ref{cpres}), are gauge invariant and cutoff independent.  

If one substitutes Eqs.\,(\ref{genvv}) into Equation~(\ref{vwti}), it follows immediately that 
\begin{equation}
\label{fsigma0}
f_{\sigma_n^H} \equiv 0\,,\; \forall H=0^{++}\, \& \,\forall n\,,
\end{equation}
because $[{\cal M}^\zeta,T^{0^{++}}]=0$; namely, for all and any states on a $0^{++}$ trajectory, whether ground-state, radial excitation or hybrid, the leptonic decay constant is zero \cite{Maris:2000ig}.  Phenomenological studies that employ the truncation of Refs.\,\cite{Munczek:1994zz,Bender:1996bb} are guaranteed to preserve this result.  Any study in which Eq.\,(\ref{fsigma0}) is not manifest violates current conservation.

This says nothing about the scalar residue, $\rho_{\sigma_n}\!(\zeta)$, knowledge of which can be useful, e.g., in constraining analyses of scalar meson production in $D$- and $B$-meson decays \cite{El-Bennich:2006yy}.  We have estimated this residue using the renormalisation-group-improved rainbow-ladder truncation of the DSEs introduced in Ref.\,\cite{Maris:1999nt}, which represents the first term in the DSE truncation mentioned above.  However, this order is not sophisticated enough to predict, e.g., the mixing angle for the $f_0$.  Hence, for illustration, the results in Table~\ref{scalarresidue} are for a pure $\bar uu$ dressed-quark state and a pure $\bar ss$ dressed-quark state.  We acknowledge that there are numerous phenomenological difficulties encountered in understanding the scalar states below $1.4\,$GeV (see, e.g., Ref.\,\cite{Pennington:2005am}) and an improved treatment might begin with an estimate of meson-loop effects via the manner employed in Ref.\,\cite{Holl:2005st}.

\begin{table}[t]
\caption{\label{scalarresidue}Ground-state scalar meson quantities calculated using the model of Ref.\,\protect\cite{Maris:1999nt} with $\omega=0.38\,$GeV and $\zeta=\zeta_{19}=19\,$GeV. NB.\ $\rho_{\sigma_n^H}(\zeta) =: m_{\sigma_n^H} \phi_{\sigma_n^H}$; $m_q \rho_{\sigma_n^H}$ is a renormalisation-point-invariant and hence one may use one- or two-loop evolution to define, e.g., $\phi_{\sigma_0^H}(\zeta=\zeta_1=1\,{\rm GeV})$.  With one-loop evolution, $\phi_{\sigma_0^{\bar u u}}(\zeta_1)=0.281\,$GeV and $\phi_{\sigma_0^{\bar s s}}(\zeta_1)=0.245\,$GeV.}
\begin{tabular}{@{}cccccc}\hline
\rule{0ex}{3ex}$H$ 
& $m_q(\zeta_{19}) $ (GeV) 
& $m_{\sigma_0^H} $ (GeV)
& $\rho_{\sigma_0^H}(\zeta_{19})$ (GeV$^2$)
& $\phi_{\sigma_0^H}(\zeta_{19})$ (GeV)
& $m_q \rho_{\sigma_0^H}$ (GeV$^3$) \\
$\bar u u$ & 0.0037 & 0.675 & (0.529)$^2$ & 0.415 & (0.101)$^3$ \\
$\bar s s$ & 0.0835 & 1.076 & (0.628)$^2$ & 0.366 & (0.321)$^3$\\\hline
\end{tabular}
\end{table}

There is a conjecture \cite{Glozman:2003bt} that DCSB notwithstanding, the properties of highly excited mesons exhibit a pattern that is consistent with a Wigner realisation of chiral symmetry.  In the present context, this could be manifested in
\begin{equation}
\rho_{\sigma_n^H}(\zeta) \approx \rho_{\pi_n^H}(\zeta)\,,\; \mbox{for}\;n\;\mbox{large}\,.
\end{equation}
$\rho_{\pi_n^H}(\zeta)$ and $\rho_{\sigma_n^H}(\zeta)$ are gauge invariant. 
Moreover these residues are different for the ground and low-lying excited $0^{-+}$ and $0^{++}$ states; e.g., $\rho_{\pi_0^{\bar uu}}(\zeta_{19})=(0.488\,{\rm GeV})^2$, $\rho_{\pi_1^{\bar uu}}(\zeta_{19})=-(0.471\,{\rm GeV})^2$ and $\rho_{\sigma_1^{\bar uu}}(\zeta_{19})=-(0.382\,{\rm GeV})^2$.  Finally, when chiral symmetry is restored at nonzero temperature, one has \cite{Maris:2000ig} $\rho_{\pi_n^H}(\zeta)=\rho_{\sigma_n^H}(\zeta)$.

The broad application of DSEs to meson properties; e.g., the illustration of these and other exact results, and the model-sensitive prediction of experimental observables, is reviewed in Ref.\,\cite{Maris:2003vk}, with many more recent studies identified by reference thereto.

\section{NUCLEONS}
\emph{Few Body 17} heard no mention of DSE baryon structure and interaction studies.  In the interim a tolerably realistic picture of nucleons has been composed.  This is not to say the study of baryons is at a level of sophistication comparable to that of mesons.  Indeed, it is at roughly the same stage as was the study of mesons more than a decade ago; namely, model-building and phenomenology constrained and informed by the best available hadron physics theory.

In quantum field theory a nucleon appears as a pole in a six-point quark Green function, with the residue proportional to the nucleon's Faddeev amplitude.  That is obtained from a Poincar\'e covariant Faddeev equation which adds-up all the possible quantum field theoretical exchanges and interactions that can take place between the three dressed-quarks which constitute the nucleon.  A merit of the Poincar\'e covariant Faddeev equation is that a modern understanding of the structure of dressed-quarks and -gluons (see Sect.\,5.1 of Ref.~\cite{Holl:2006ni}) is straightforwardly incorporated; viz., effects owing to and arising from the strong momentum dependence of these propagators are realised and exhibited. 

The tractable treatment of the Faddeev equation requires a truncation.  One is founded \cite{Cahill:1988dx} on the observation that an interaction which describes colour-singlet mesons also generates quark-quark (diquark) correlations in the colour-$\bar 3$ (antitriplet) channel \cite{Cahill:1987qr}.  The dominant correlations for ground state octet and decuplet baryons are scalar and axial-vector diquarks.  This can be understood on the grounds that: the associated mass-scales are smaller than the baryons' masses \cite{Burden:1996nh,Maris:2002yu}, with models typically giving (in GeV)
$m_{[ud]_{0^+}} = 0.74 - 0.82$,
$m_{(uu)_{1^+}}=m_{(ud)_{1^+}}=m_{(dd)_{1^+}}=0.95 - 1.02$;
the electromagnetic size of these correlations is less than that of the proton \cite{Maris:2004bp} -- $r_{[ud]_{0^+}} \approx 0.7\,{\rm fm}$, from which one may estimate $r_{(ud)_{1^+}} \sim 0.8\,{\rm fm}$ based on the $\rho$-meson/$\pi$-meson radius-ratio \cite{Burden:1995ve,Hawes:1998bz}; and the positive parity of the correlations matches that of the baryons.  Both scalar and axial-vector diquarks provide attraction in the Faddeev equation.

\begin{figure}[t]
\centerline{
\includegraphics[width=0.60\textwidth]{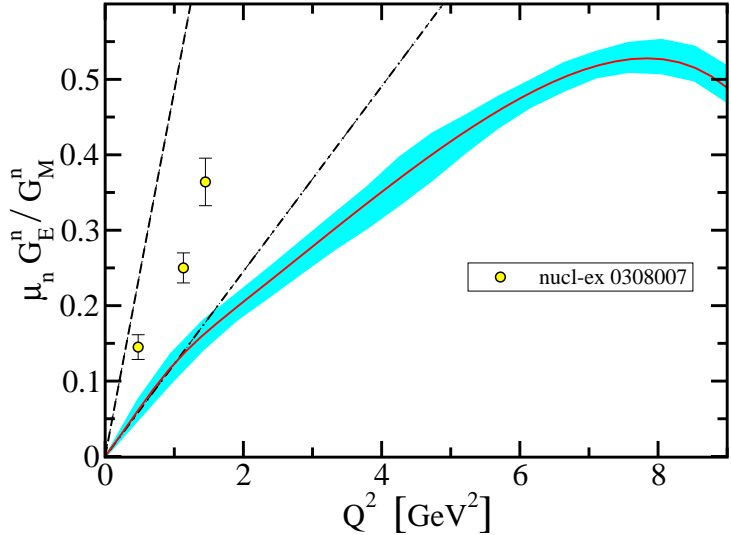}}
\vspace*{-4ex}

\caption{\label{GEnGMn}\emph{Solid curve} -- $\mu_n\, G_E^n(Q^2)/G_M^n(Q^2)$ obtained from the form factors calculated in Ref.\,\protect\cite{Alkofer:2004yf,Holl:2005zi}.  The band shows the response of the ratio to changes in the axial-vector diquark's electromagnetic properties; viz., variation of $\pm 1$ about the reference values: $\mu_{1^+}=2$, $\chi_{1^+}=1$ and $\kappa_{\cal T} = 2$, and also reflects Monte-Carlo-integration error.  \emph{Dashed curve} -- $-(1/6) (r_n^2)^{\rm exp.} Q^2$; \emph{dash-dot curve} -- $-(1/6) (r_n^2)^{\rm calc.} Q^2$. Data from Ref.\,\protect\cite{Madey:2003av}.}  
\end{figure}

The truncation of the Faddeev equation's kernel is completed by specifying that the quarks are dressed, with two of the three dressed-quarks correlated always as a colour-$\bar 3$ diquark.  Binding is then effected by the iterated exchange of roles between the bystander and diquark-participant quarks.  This ensures that the Faddeev amplitude exhibits the correct symmetry properties under fermion interchange.  A Ward-Takahashi-identity-preserving electromagnetic current for the baryon thus constituted is subsequently derived~\cite{Oettel:1999gc}.  It depends on the electromagnetic properties of the axial-vector diquark correlation: its magnetic and quadrupole moments, $\mu_{1^+}$ and $\chi_{1^+}$, respectively; and the strength of electromagnetically induced axial-vector $\leftrightarrow$ scalar diquark transitions, $\kappa_{\cal T}$.

Figure~\ref{GEnGMn} depicts the calculated ratio $\mu_n G_E^n(Q^2)/G_M^n(Q^2)$.  The behaviour of the analogous proton ratio was published in Ref.\,\cite{Alkofer:2004yf,Holl:2005zi}: it passes through zero at $Q^2\approx 6.5\,$GeV$^2$.  It is noteworthy that in the neighbourhood of $Q^2=0$, 
\begin{equation}
\label{smallQ}
\mu_p\,\frac{ G_E^n(Q^2)}{G_M^n(Q^2)} = - \frac{r_n^2}{6}\, Q^2 ,
\end{equation}
where $r_n$ is the neutron's electric radius.  Our calculation shows this to be a good approximation for $r_n^2 Q^2 \nlsim 1$.  The data \cite{Madey:2003av} are consistent with this.  The calculated curve omitted pion-loop effects from the current and therefore underestimated the magnitude of $r_n^2$.   This is apparent in the figure.  It is thus evident that, just as for the proton, the small $Q^2$ behaviour of this ratio is materially affected by the neutron's pion cloud.  

Pseudoscalar mesons are not pointlike and therefore their contributions to form factors diminish in magnitude with increasing $Q^2$ (see, e.g., Refs.~\cite{Alkofer:1993gu,Sato:2000jf,Miller:2002ig,Hammer:2003qv}).  It follows therefore that the evolution of $\mu_n G_E^n(Q^2)/G_M^n(Q^2)$ on $Q^2\ngsim 2\,$GeV$^2$ is primarily determined by the quark-core of the neutron.  This calculation predicts that the ratio will continue to increase steadily until $Q^2\simeq 8\,$GeV$^2$.

In a Poincar\'e covariant treatment the quark core of a nucleon is necessarily described by a Faddeev amplitude with nonzero quark orbital angular momentum.  This is why the Faddeev amplitude is a matrix-valued function with a rich structure that, in a baryons' rest frame, corresponds to a relativistic wave function with $s$-wave, $p$-wave and even $d$-wave components~\cite{Oettel:1998bk}.  Figure~\ref{GEnGMn} illustrates that while there is some quantitative sensitivity to the electromagnetic structure of the diquark correlations, the gross features of the form factors are primarily governed by correlations expressed in the nucleon's Faddeev amplitude and, in particular, by the amount of intrinsic quark orbital angular momentum \cite{Bloch:2003vn}.  The nature of the kernel in the Faddeev equation specifies just how much quark orbital angular momentum is present in a baryon's rest frame. 

\section{CLOSING}
Two emergent phenomena are responsible for the observed properties of hadrons: confinement and dynamical chiral symmetry breaking.  They can be viewed as an essential consequence of the presence and role of particle-antiparticle pairs in an asymptotically free theory and therefore can only be veraciously understood in relativistic quantum field theory.  Quantum mechanics is inadequate and supposed successes within that framework should be viewed as a consequence of fine-tuning or good fortune.  

DSEs are well suited to the study of hadron physics at and beyond the frontiers being explored by modern, high-luminosity facilities.  The framework quite naturally wraps each of QCD's elementary excitations in a cloud of virtual particles that is exceedingly dense at low momentum.  It admits a systematic, symmetry preserving and nonperturbative truncation scheme, and thereby gives access to continuum strong QCD.  The tractability of the truncation scheme makes available a practical tool for the prediction of observables.  The consequent opportunities for rapid feedback between experiment and theory brings within reach an intuitive understanding of strong interaction phenomena.  Furthermore, a dialogue between DSE and lattice-QCD studies is today proving fruitful.

A pressing task is the drawing of an accurate map of the confinement force between light-quarks within hadrons.  The careful and quantitative study of mesons in the mass range $1$ -- $2\,$GeV appears to us as a certain way to make progress in this direction.

\bibliographystyle{elsart-num}
\bibliography{CDRobertsFB18}

\end{document}